\documentclass[prb,twocolumn,showpacs]{revtex4}
\usepackage{graphicx}
\usepackage{amsmath}
\usepackage{amssymb}
\usepackage{dcolumn}
\usepackage{float}
\usepackage{bm}
\usepackage[breaklinks=true,colorlinks,citecolor=blue,linkcolor=blue,urlcolor=blue]{hyperref}

\renewcommand{\i}{\mathrm{i}}

\DeclareMathAlphabet{\bi}{OML}{cmm}{b}{it}

\def\be{\begin{equation}}
\def\ee{\end{equation}}
\def\bearr{\begin{eqnarray}}
\def\eearr{\end{eqnarray}}
\def\la{\langle}
\def\ra{\rangle}

\begin{document}
\title{Electron-phonon interaction in a spin-orbit coupled
quantum wire with a gap}
\bigskip

\author{Tutul Biswas and Tarun Kanti Ghosh}
\normalsize
\affiliation
{Department of Physics, Indian Institute of Technology-Kanpur,
Kanpur-208 016, India}
\date{\today}
 
\begin{abstract}
Interaction between electron and acoustic phonon in an in-plane magnetic field induced 
gapped quantum wire with Rashba spin-orbit interaction is studied.
We calculate acoustic phonon limited resistivity ($\rho$) and phonon-drag 
thermopower ($S_g$) due to two well known mechanisms of electron-phonon
interaction namely, deformation potential (DP) and piezoelectric (PE)
scattering. In the so called Bloch-Gruneisen temperature limit both
$\rho$ and $S_g$ depend on temperature ($T$) in a power law fashion
i.e. $\rho$ or $S_g\sim T^{\nu_T}$. For resistivity, $\nu_T$ takes the value
$5$ and $3$ due to DP and PE scattering respectively. On the 
other hand, $\nu_T$ is $4$ and $2$ due to DP and PE scattering,
respectively for phonon-drag thermopower. Additionally, we find numerically
that $\nu_T$ depends on Rashba parameter ($\alpha$) and electron density ($n$). The dependence of $\nu_T$ 
on $\alpha$ becomes more prominent at lower density. We also study 
the variations of $\rho$ and $S_g$ with carrier density in the Bloch-Gruneisen
regime. Through a numerical analysis a similar power law dependence $\rho$ or $S_g\sim n^{-\nu_n}$ is 
established in which the effective exponent $\nu_n$ undergoes a smooth transition from a 
low density behavior to a high density behavior. At a higher density regime, 
$\nu_n$ matches excellently with the value obtained from theoretical arguments.
Approximate analytical expressions for both
resistivity and phonon-drag thermopower in the Bloch-Gruneisen regime are given.

\end{abstract}


\maketitle

 \section{Introduction}

Due to the promising applications in the area of
quantum information processing\cite{inform} and device technology,\cite{device}
spin dependent phenomena\cite{spin1,spin2,spin3} in low dimensional structures 
have been of major interest in scientific communities 
for several years. The main route of spin related 
phenomena is the well known spin-orbit interaction (SOI). 
In semiconductor structures, SOI originates due to the 
inversion symmetry breaking either in the bulk or at the hetero-interface.
Band bending in heterostructure gives rise to an electric 
field to produce an asymmetric confining potential which
itself is responsible for generating a SOI of Rashba type.\cite{rashba} The strength of 
Rashba SOI is proportional to the magnitude of the 
electric field generated and hence is tunable\cite{tune1,tune2} with
the aid of an external gate voltage.
Another kind of SOI, usually termed as Dresselhaus SOI,\cite{Dressh} 
originates due to the breaking of inversion symmetry in the 
bulk crystal. In general the strength\cite{strng_RD} of the Dresselhaus
SOI is smaller than that of Rashba SOI in heterostructure.
The case of equal strength of both SOIs is of particular 
importance for future scope of developing non-ballistic spin field effect
transistor.\cite{NBspin_FET}

In a quantum well, restriction of carrier's motion 
by an additional confinement in a particular direction  essentially leads to the formation
of a quantum wire (QW). The width of QW is of the order of the Fermi
wave length in order to allow ballistic transport.\cite{BLT_QW}
Semiconductor QW with SOI is considered as a building block for a 
better implementation\cite{spinFT} of spin-field effect transistor.\cite{Datta_Das}
An in-plane magnetic field along the wire direction lifts the $k=0$ degeneracy in a spin-orbit coupled QW
and as a result a gap is induced in the energy spectrum.
Immense interests have been grown on the gapped spin-orbit coupled QW 
because of several proposals of asymmetric spin filtering,\cite{helix1}
controlling impurities\cite{mag_imp} by a magnetic field,
topological superconducting phase,\cite{topo_ph1,topo_ph2,topo_ph3}
helical states\cite{helix1,topo_ph2,helix2,helix3} etc. Recently, magnetic field induced
spin-orbit gap in an one dimensional hole gas has been realized 
experimentally.\cite{helix3}

In the present study we mainly focus on various consequences 
of electron-phonon interaction in a spin-orbit coupled gapped QW
by calculating phonon limited resistivity and phonon-drag thermopower
in the Bloch-Gruneisen (BG) regime. In the BG regime, the resistivity
differs abruptly from its equipartition behavior. An upper bound of the 
BG regime can be defined by the characteristic temperature 
$T_{BG}=2\hbar v_s k_F/k_B$, where $v_s$ is the sound velocity and 
$k_F$ is the Fermi wave vector. Below $T_{BG}$ the acoustic phonon 
energy is comparable with the thermal energy. Due to the smallness of 
Fermi surface in semiconductor structures, acoustic phonon with
wave vector $q\simeq 2k_F$ can not be excited appreciably which in 
turn leads to a complicated temperature dependence of resistivity 
below $T_{BG}$. The existence of the BG regime in semiconductor quantum well
has been confirmed experimentally.\cite{BG_obs} Another important
quantity that can be used to probe 
the electron-phonon interaction in semiconductor nanostructure is the 
phonon-drag contribution to the thermoelectric power. A number studies 
have been performed to understand the behavior of phonon limited 
mobility\cite{ph_mob1,ph_mob2,ph_mob3,ph_mob1d1,ph_mob1d2,ph_mob1d3,ph_mob1d4,ph_mob1d5} and
phonon-drag thermopower\cite{phdrg1,phdrg2,phdrg3,phdrg4,phdrg5}
in quantum well and wire for several years. Recently, acoustic phonon
limited resistivity\cite{tutul1} and phonon-drag thermopower\cite{tutul2} in a Rashba spin-orbit
coupled two dimensional electron gas have been studied. It is revealed 
through a numerical analysis that the effective exponents of the temperature
dependence of both resistivity and thermopower depend significantly on the 
strength of the Rashba SOI.

In BG regime, we find analytically that both resistivity and phonon-drag thermopower 
in a spin-orbit coupled gapped QW maintains a power law dependence with
temperature i.e. $\rho$ or $S_g\sim T^{\nu_T}$. The effective exponent $\nu_T$
is $5$ and $3$ due to deformation potential  (DP) and piezoelectric (PE) scattering,
respectively in the case of phonon limited resistivity.
For phonon-drag thermopower $\nu_T$ becomes $4$ and $2$ due to DP and PE 
scattering, respectively. The exponent $\nu_T$ has also been extracted numerically
which clearly undergoes a transition from BG regime to equipartition limit.
At a relatively higher density, $\nu_T$ is in excellent agreement with the 
analytical results. Temperature variation of $\nu_T$ for different 
Rashba parameter ($\alpha$) at a fixed density has also been shown.
The effect of $\alpha$ on the temperature dependence of $\nu_T$
becomes less prominent as one approaches towards higher density. 
Additionally, the dependence of both $\rho$ and $S_g$ on carrier 
density has been shown in which both quantities undergo a
transition from a relatively low to a higher density behavior.

We organize this paper in the following way. 
In section II we present all the theoretical details.
Numerical results and discussions have been reported 
in section III. We summarize our work in section IV.

\section{Theory}

\subsection{Physical system}

We consider a semiconductor QW of radius $R$ in which electrons are free
to move in the $z$- direction. Rashba SOI in a QW essentially breaks the spin-degeneracy\cite{zulicke} 
by shifting the non-degenerate subbands laterally along the wave vector. But the spectrum 
is still degenerate at $k=0$. An external magnetic field ${\bf B}=B\hat{z}$ 
along the wire direction can be used to lift this degeneracy further 
by inducing a gap $\Delta$ in the energy spectrum.\cite{helix1,privman,meden}
Now the single particle Hamiltonian describing 
a gapped QW with Rashba SOI is written as
\begin{eqnarray}\label{Hamil}
H=\Big(\frac{p^2}{2m^\ast}+V({\bf r})\Big)\sigma_0+\frac{\alpha}{\hbar}\sigma_yp
+\Delta\sigma_z,
\end{eqnarray}
where $p=\hbar k$ is the electron momentum in the $z$- direction, $m^\ast$ is effective mass of
electron, $\sigma_0$ is the unit matrix, 
$\sigma_i$'s are the Pauli spin matrices, and $\alpha$ is
the strength of RSOI. Also $\Delta=g^\ast\mu_B B/2$ is the Zeeman 
energy with $g^\ast$ and $\mu_B$ as the effective Lande $g$- factor and Bohr magneton,
respectively. Finally, $V({\bf r})$ is the confining potential 
in the transverse direction ${\bf r}\equiv(x,y)$.

The wire is assumed to be thin enough so that only the 
lowest subband in the transverse direction is occupied 
by electrons. By diagonalizing Eq. (\ref{Hamil}),
the eigen energies corresponding to the present physical system can be found in the following form 
\begin{eqnarray}\label{eig_eng}
\epsilon_k^\lambda=\frac{\hbar^2 k^2}{2m^\ast} + \lambda
\sqrt{\alpha^2 k^2+\Delta^2},
\end{eqnarray}
where $\lambda=\pm$ describes the branch index. Note that
the energy is measured from the bottom of the lowest sub-band 
energy $\varepsilon_{10}=\hbar^2 k_{10}^2/(2m^\ast)$ with 
$k_{10}$ as the sub-band wave vector.
 
The eigen functions corresponding to the $+$ and $-$ branches are respectively
given by
\begin{eqnarray}
 \Psi^{+}(r,z)=\frac{e^{ikz}}{\sqrt{2\pi}}
\begin{pmatrix}
 \cos{\frac{\phi_k}{2}}\\i\sin{\frac{\phi_k}{2}}
\end{pmatrix}
\Phi_{10}(r),
\end{eqnarray}

and

\begin{eqnarray}
 \Psi^{-}(r,z)=\frac{e^{ikz}}{\sqrt{2\pi}}
\begin{pmatrix}
 \sin{\frac{\phi_k}{2}}\\-i\cos{\frac{\phi_k}{2}}
\end{pmatrix}
\Phi_{10}(r),
\end{eqnarray}
where $\tan{\phi_k}=\alpha k/\Delta$. The lowest subband
wave function for $r\leq R$ is given by $\Phi_{10}(r)=J_0(k_{10}r)/\{\sqrt{\pi R^2}J_1(k_{10}R)\}$
with $J_\nu(x)$ as the Bessel function of order $\nu$. It is important to mention
that $k_{10}R$ is the first zero of $J_0(k_{10}R)$. Out side the quantum wire $\Phi_{10}(r)$
vanishes.

At a fixed energy namely, Fermi energy $\epsilon_F$, one can have the following 
expression for the Fermi wave vectors
\begin{eqnarray}\label{ferm_wave}
k_F^\lambda=\Bigg\{\bigg(-\lambda k_\alpha+\sqrt{k_\alpha^2+
\frac{2m^\ast\epsilon_F}{\hbar^2}+\frac{\Delta^2}{\alpha^2}}\bigg)^2-\frac{\Delta^2}{\alpha^2}\Bigg\}^\frac{1}{2},
\end{eqnarray}
where $k_\alpha=m^\ast\alpha/\hbar^2$.
In the $B\rightarrow 0$ limit, Eq. (\ref{ferm_wave}) reduces to the known forms of the Fermi wave vectors 
of a Rashba spin-orbit coupled quantum wire.

The velocity corresponding to the energy spectrum given in Eq. (\ref{eig_eng}) is 
calculated as 
\begin{eqnarray}\label{vel}
 v_k^\lambda=\frac{\hbar k}{m^\ast} + \lambda 
\frac{\alpha^2 k}{\hbar \sqrt{\alpha^2k^2+\Delta^2}}.
\end{eqnarray}

\subsection{Phonon limited resistivity}
In this section we shall calculate resistivity 
due to the electron-phonon interaction using 
Boltzmann transport theory. We restrict ourselves 
to consider only the longitudinal and transverse acoustic 
phonon modes. Using Drude's formula, 
the resistivity is simply written as
\begin{eqnarray}
\rho=\frac{m^\ast}{ne^2}\Big \la \frac{1}{\tau} \Big\ra,
\end{eqnarray}
where $\la 1/\tau \ra$ is the inverse relaxation time (IRT) averaged over 
energy and $n$ is the density of electron.

The energy averaged IRT for a specific energy branch $\lambda$  is given by
\begin{eqnarray}
\Big \la \frac{1}{\tau^\lambda}\Big \ra=\frac{2}{k_BT} \int d\epsilon_k^\lambda
f(\epsilon_k^\lambda)\{1-f(\epsilon_k^\lambda)\}
\frac{1}{\tau(\epsilon_k^\lambda)},
\end{eqnarray}
where $f(\epsilon_k^\lambda)=[1+e^{\beta(\epsilon_k^\lambda-\epsilon_F)}]^{-1}$ is the Fermi-Dirac
distribution function with $\beta=(k_BT)^{-1}$. Here $2$- factor appears to consider
the $k<0$ contribution since the energy spectrum is symmetric about $k=0$.

According to the Boltzmann transport theory, the  IRT can be 
found in the following semi-classical form
\begin{eqnarray}\label{invRT}
\frac{1}{\tau(\epsilon_k^\lambda)}=\sum_{ k^\prime, \lambda^\prime}
\Big(1-\frac{k^\prime}{k}\Big) W_{k,k^\prime}^{\lambda, \lambda^\prime}
\frac{1-f(\epsilon_{k^\prime}^{\lambda^\prime})}{1-f(\epsilon_{k}^\lambda)}.
\end{eqnarray}

The transition probability from an initial state $\vert k, \lambda \ra$ to 
a final state $\vert k^\prime, \lambda^\prime \ra$ is given by 
the Fermi's Golden rule as

\begin{widetext}

\begin{eqnarray}\label{trans_prob}
 W_{k,k^\prime}^{\lambda,\lambda^\prime}=\frac{2\pi}{\hbar}\sum_{\bf Q}
\vert C_Q \vert^2 \vert F(q_\bot)\vert^2
\vert \xi_{k,k^\prime}^{\lambda,\lambda^\prime} \vert^2
\Big\{N_Q \delta\big(\epsilon_{k^\prime}^{\lambda^\prime}-\epsilon_k^\lambda
-\hbar \omega_Q\big) \delta_{k^\prime,k+q}
+\big(N_Q+1\big) \delta\big(\epsilon_{k^\prime}^{\lambda^\prime}-\epsilon_k^\lambda
+\hbar \omega_Q\big) \delta_{k^\prime,k-q}\Big\},
\end{eqnarray}

\end{widetext}
where ${\bf Q}=(q_\bot,q)$ is the phonon wave vector with 
$q_{\bot}=(q_x,q_y)$ and $q=q_z$,
$N_Q=[e^{\beta\hbar\omega_Q}-1]^{-1}$ is the phonon distribution function, 
$C_Q$ is the matrix element corresponding to the electron-phonon interaction,
$F(q_\bot)$ is the form factor arising due to the transverse confinement. The first and second terms 
in the braces of Eq. (\ref{trans_prob}) correspond to the absorption and emission of acoustic phonons,
respectively. Finally, the overlap integral 
$\vert \xi_{k,k^\prime}^{\lambda,\lambda^\prime} \vert^2$ coming from the spinor part 
of the wave function is given by
\begin{eqnarray}\label{spinor_over}
\vert \xi _{k,k^\prime}^{\lambda,\lambda^\prime} \vert^2=
\frac{1+\lambda \lambda^\prime \cos(\phi_{k^\prime}-\phi_k)}{2}.
\end{eqnarray}

The square of the electron-phonon matrix elements corresponding to DP and  PE scatterings
are respectively given by 

\begin{eqnarray}\label{mat_DP}
\vert C_Q \vert^2=\frac{D^2\hbar Q}{2\rho_m v_{sl}},
\end{eqnarray}

\begin{eqnarray}\label{mat_PE}
\vert C_{Q,l(t)} \vert^2=
\frac{(eh_{14})^2\hbar}{2\rho_m v_{sl(t)}}
\frac{A_{l(t)}}{Q},
\end{eqnarray}
where $D$ is deformation potential strength, $h_{14}$ is the 
relevant PE tensor component, $\rho_m$ is the mass density, and
$v_{sl(t)}$ is the longitudinal (transverse) component of 
sound velocity. The anisotropic factors are given by 
$A_{l}=9q^2q_{\bot}^4/(2Q^6)$ and 
$A_t=(8q^4q_{\bot}^2+q_{\bot}^6)/(4Q^6)$.
 
Finally, the square of the form factor is defined as
\begin{eqnarray}
\vert F(q_\bot)\vert^2=\Big\vert \int \Phi_{10}^\ast({\bf r})
e^{i{\bf q_\bot}\cdot{\bf r}} \Phi_{10}({\bf r})  d{\bf r} \Big\vert^2.
\end{eqnarray}

In literature, it is assumed\cite{Lee,Fishman} that the 
electron density is sufficiently high near the axis of the 
wire and vanishes everywhere and consequently one replaces
$\vert \Phi_{10}({\bf r})\vert^2\sim 1/(\pi R_0^2)$, where $R_0< R$ 
defines some confinement region. So in
this approximation the square of the form factor can be 
readily obtained in the following exact form as
$ \vert F(q_\bot)\vert^2=4\{ J_1(q_\bot R_0)/(q_\bot R_0)\}^2 $.
In the rest of the paper we will be using this expression of the
form factor.

Let us now discuss the possibility of intra
and inter-branch scatterings. Generally, at low temperature intra-branch
scatterings are dominant. To occur inter-branch scattering large momentum
transfer is needed. Since in the BG regime $q\ll k_F$ so the possibility of
inter-branch scatterings are ruled out. In a spin-orbit coupled QW without
the gap, the inter-branch scattering
is strictly forbidden as readily understood from Eq. (\ref{spinor_over})
since $\cos(\phi_k-\phi_{k^\prime})$ becomes unity as $B\rightarrow 0$.
But for $B\neq 0$ inter-branch scattering are possible. We have checked
numerically that the inter-branch contribution is very small in comparison
with the intra-branch one. So henceforth we will consider only
the intra-branch scattering.

Now, the summation over $k^\prime$ in Eq. (\ref{invRT}) can be performed
with the aid of  $\delta_{k^\prime,k\pm q}$  given in Eq. (\ref{trans_prob}).
In the BG regime, phonon energy is comparable with the thermal 
energy but much less than the Fermi energy i.e.
$\hbar \omega_Q\simeq k_BT \ll \epsilon_F$. So one can safely make the 
following approximation $f(\epsilon_k)\{1-f(\epsilon_k \pm \hbar\omega_Q)\}
\simeq \hbar \omega_Q (N_Q+1/2 \pm 1/2)\delta(\epsilon_k-\epsilon_F)$, 
where $+$ and $-$ signs correspond to the absorption and emission of acoustic 
phonons, respectively. Using the above mentioned simplification and 
approximation one can find the energy averaged IRT in the following form
\begin{eqnarray}\label{IRT_av}
\Big \la \frac{1}{\tau^\lambda} \Big \ra&=&\frac{4\pi}{k_B T}\sum_{{\bf Q}}
\vert C_Q \vert^2 \vert F(q_\bot) \vert^2  \frac{q\omega_Q}{k_F^\lambda}
N_Q(N_Q+1)\nonumber\\
&\times& \Bigg\{ \vert \xi_{k_F,k_F-q}^{\lambda,\lambda} \vert^2 
\delta\big(\epsilon_{k_F-q}^{\lambda}-\epsilon_{k_F}^\lambda
+\hbar \omega_Q\big)\nonumber\\
&-&\vert \xi_{k_F,k_F+q}^{\lambda,\lambda} \vert^2 
\delta\big(\epsilon_{k_F+q}^{\lambda}-\epsilon_{k_F}^\lambda
-\hbar \omega_Q\big)\Bigg\}.
\end{eqnarray}

The summation over ${\bf Q}$ in Eq. (\ref{IRT_av}) can be transformed into an integration over
$q$ and $q_\bot$ as $\sum_{\bf Q}\rightarrow(1/4\pi^2)\int q_\bot dq_\bot dq$. At very low 
temperature (BG regime) phonon states with wave vector $q\ll k_F$ are populated.
The delta functions given in Eq. (\ref{IRT_av}) can be approximated 
in the following form as (see Appendix A for details)
\begin{eqnarray}\label{delta_fn}
\delta(\epsilon_{k_F\pm q}^\lambda-\epsilon_{k_F}^\lambda \mp \hbar\omega_Q)
&\simeq& \frac{m^\ast}{\hbar^2 \tilde{k}_F^\lambda}
\bigg\{1\mp \frac{m^\ast v_s}{\hbar \tilde{k}_F^{\lambda^2}}
\tilde{g}_\alpha^\lambda q_\bot\bigg\} \nonumber\\
&\times& \delta \Big(q-\frac{m^\ast v_s}{\hbar \tilde{k}_F^\lambda}q_{\bot}\Big),
\end{eqnarray}
where $\tilde{k}_F^\lambda=k_F^\lambda
(1+2\lambda \varepsilon_\alpha/\varepsilon_{k_F^\lambda})$ with 
$\varepsilon_{k_F^\lambda}=\sqrt{\alpha^2 k_F^{\lambda^2}+\Delta^2}$, 
$\varepsilon_\alpha=m^\ast\alpha^2/(2\hbar^2)$
and $\tilde{g}_\alpha^\lambda=1+2\lambda \tilde{\varepsilon}_\alpha/\varepsilon_{k_F^\lambda}$
with $\tilde{\varepsilon}_\alpha = m^\ast \tilde{\alpha}^2/(2\hbar^2)$. Here 
$\tilde{\alpha}$ is defined as $\tilde{\alpha}=\alpha \Delta/\varepsilon_{k_F^\lambda}$.

Now inserting Eq. (\ref{delta_fn}) into Eq. (\ref{IRT_av})
and putting the expressions for $\vert \xi_{k_F,k_F\pm q}^{\lambda,\lambda} \vert^2$
we have 
\begin{eqnarray}\label{IRT_fn}
\Big \la \frac{1}{\tau^\lambda}\Big \ra&=&\frac{m^\ast v_s}{\pi \hbar^2 k_B T}
\frac{1}{k_F^\lambda \tilde{k}_F^\lambda} \int dq dq_\bot \vert C_Q \vert^2
\vert F(q_\bot) \vert^2 q q_\bot Q \nonumber\\
&\times& N_Q (N_Q+1)\delta\Big(q-\frac{m^\ast v_s}{\hbar \tilde{k}_F^\lambda}q_{\bot}\Big)
\Bigg\{ \frac{\cos{\phi_{k_F}^-} -\cos{\phi_{k_F}^+}}{2} \nonumber\\
&+& \frac{m^\ast v_s}{\hbar \tilde{k}_F^{\lambda^2}}
\tilde{g}_\alpha^\lambda q_\bot
 \Big[1+\frac{\cos{\phi_{k_F}^-} +\cos{\phi_{k_F}^+}}{2}\Big]\Bigg\},
\end{eqnarray}
where $\phi_{k_F}^\pm=\phi_{k_F}-\phi_{k_F \pm q}$.

\subsection{Phonon-drag thermopower}
In presence of a temperature gradient diffusion motion of electron 
takes place. As a result of electron-phonon interaction, phonon gains a 
finite heat flux which in turn drags electron along with it from the 
hot to the cold end. In this way a phonon-drag contribution to the thermoelectric power is 
generated. Phonon-drag thermopower is more fundamental 
quantity in probing electron-phonon strength experimentally.

To calculate phonon-drag thermopower two approaches
named as $Q$ and $\Pi$-approach are mainly followed.
In the rest we follow only the $Q$-approach. 
We now start with the following expression\cite{phdrg1} for the phonon-drag thermopower 
\begin{eqnarray} \label{phdrag1}
S_g^\lambda&=&\frac{e\tau_p}{\sigma Lk_BT^2} 
\sum_{\lambda^\prime}\sum_{{\bf k}, 
{\bf k^\prime}, {\bf Q}} \hbar\omega_Q f(\epsilon_k^\lambda)
\Big[1-f(\epsilon_{k^\prime}^{\lambda^\prime})\Big]\nonumber\\
& \times & W_{Q,\rm Ab}^{\lambda\lambda^\prime}({\bf k},{\bf k^\prime}) 
\Big\{\tau(\epsilon_k^\lambda){\bf v}_k^\lambda-
\tau(\epsilon_{k^\prime}^{\lambda^\prime})
{\bf v}_{k^\prime}^{\lambda^\prime}\Big\}\cdot{\bf v}_p,
\end{eqnarray}
where  $\tau_p$ is the phonon 
mean free time, $L$ is the length of the sample, 
$\sigma$ is the Drude conductivity, 
 $\tau(\epsilon_k)$ is electron's
momentum relaxation time, 
${\bf v}_k^\lambda$ is the velocity of an electron
as given in Eq. (\ref{vel}), ${\bf v}_p=v_s \hat{Q}$ is the 
velocity of phonon. Finally,
$W_{Q, \rm Ab}^{\lambda\lambda^\prime}({\bf k},{\bf k^\prime})$ is the transition 
probability by which an electron makes a transition
from an initial state $\vert{\bf k},\lambda\ra$ to a final state 
$\vert{\bf k^\prime},\lambda^\prime\ra$ with the absorption of an acoustic phonon.

The transition probability can be 
written as

\begin{eqnarray}\label{trans_rate}
W_{QAb}^{\lambda\lambda^\prime}&=&\frac{2\pi}{\hbar}
\vert C_ Q\vert^2 \vert F(q_\bot)\vert^2 
\vert \xi_{k,k^\prime}^{\lambda,\lambda^\prime} \vert^2 N_{Q}
\delta\Big(\epsilon_{k^\prime}^{\lambda^\prime} - 
\epsilon_k^\lambda-\hbar\omega_Q\Big)\nonumber\\
&\times&\delta_{k^\prime,{ k+ q}}.
\end{eqnarray}

In Eq. (\ref{phdrag1}) summation over $k^\prime$ is readily done 
using $\delta_{k^\prime,k+q}$ given in Eq. (\ref{trans_rate}). Further a
slow variation of $\tau(\epsilon_{k})$ over an energy scale $\sim \hbar\omega_Q$ is
assumed. So one can use the approximation
$\tau(\epsilon_{k}+\hbar\omega_Q)\simeq\tau(\epsilon_{k})$.

The summation over ${\bf k}$ in Eq. (\ref{phdrag1}) can be converted into
an integral over $\epsilon_{k}$ by the following transformation
\begin{eqnarray}\label{trans}
\sum_{\bf k}\rightarrow\frac{m^\ast L}{2\pi \hbar^2}
\int \frac{1}{k^\lambda}\Bigg(1-\lambda 
\frac{k_\alpha}{\sqrt{k_\alpha^2+\frac{2m^\ast \epsilon_k}{\hbar^2}+\frac{\Delta^2}{\alpha^2}}}
\Bigg) d\epsilon_k,
\end{eqnarray}
where $k^\lambda$ can be obtained from Eq. (\ref{ferm_wave}).

Using Eq. (\ref{phdrag1}-\ref{trans}) one can finally obtain 
the following expression for the phonon-drag thermopower for a specific
branch $\lambda$ in the BG regime as

\begin{eqnarray}\label{phdrag}
S_g^\lambda&=&-\frac{{m^\ast}^3 l_p v_s}{4\pi^2 n e \hbar^3 k_B T^2}
\frac{N_F^\lambda}{k_F^\lambda\tilde{k}_F^\lambda} 
\int dq dq_\bot \vert C_Q \vert^2 \vert F(q_\bot) \vert^2 q_\bot \nonumber\\
&\times& Q^2 N_Q(N_Q+1)\Big(1-\frac{m^\ast v_s}{\hbar \tilde{k}_F^{\lambda^2}}
\tilde{g}_\alpha^\lambda q_\bot \Big)
\delta\Big(q-\frac{m^\ast v_s}{\hbar \tilde{k}_F^\lambda}q_{\bot}\Big)\nonumber\\
&\times& \vert \xi_{k_F,k_F+q}^{\lambda,\lambda} \vert^2
\Big[{\bf v}_{k_F+q}^\lambda - {\bf v}_{k_F}^\lambda \Big]\cdot {\bf v}_p,
\end{eqnarray}
where $N_F^\lambda=\Big(1-\lambda k_\alpha/\sqrt{k_\alpha^2+
\frac{2m^\ast \epsilon_F}{\hbar^2}+\frac{\Delta^2}{\alpha^2}}\Big)$.
Note that in deriving Eq. (\ref{phdrag}) we have used the approximation 
$f(\epsilon_k)\{1-f(\epsilon_k + \hbar\omega_Q)\}
\simeq \hbar \omega_Q (N_Q+1)\delta(\epsilon_k-\epsilon_F)$ as earlier.

\subsection{Approximate analytical results in BG regime}

We shall now derive some approximate analytical expressions for phonon 
limited resistivity and phonon-drag thermopower in the BG regime.

In the BG regime we have $q<<k_F$. In this limit one can use 
following approximation $\cos{\phi_{k_F}^\pm}\simeq 1$.
Again phonon energy is comparable with the thermal energy in 
the BG regime i.e. $\hbar v_s q_\bot\sim k_BT$. So we have 
$q_\bot R_0=k_B T R_0/\hbar v_s $ which is much lower than unity 
and consequently one can approximate\cite{phdrg4} the form factor as
$\vert F(q_\bot)\vert^2\simeq 1$.

Under the above mentioned approximations, Eq. (\ref{IRT_fn}) takes 
the following form as 
\begin{eqnarray}\label{IRT_aprx}
\Big \la \frac{1}{\tau^\lambda}\Big \ra&\simeq&\frac{2}{\pi \hbar k_B T}
\Big(\frac{m^\ast v_{s}}{\hbar \tilde{k}_F^\lambda}\Big)^3
\frac{\tilde{g}_\alpha^\lambda}{k_F^\lambda \tilde{k}_F^\lambda}\nonumber\\
&\times&\int dq_\bot q_\bot^4 \vert C_{q_\bot} \vert^2 N_{q_\bot}(N_{q_\bot}+1).
\end{eqnarray}

Now inserting $\vert C_{q_\bot}\vert^2$ as given in 
Eq. (\ref{mat_DP}-\ref{mat_PE}) and using the standard integral 
$\int_0^\infty x^p e^x/(e^x-1)^2=\zeta(p) p!$ in Eq. (\ref{IRT_aprx}) one can derive 
the following expressions for the energy averaged IRT corresponding to 
DP, longitudinal and transverse PE scatterings, respectively as

\begin{eqnarray}\label{apprx_DP}
\Big \la \frac{1}{\tau^\lambda}\Big \ra_{\rm DP} \simeq 
\frac{D^2 \tilde{g}_\alpha^\lambda 5! \zeta(5)}{\pi \rho_m \hbar^6 v_{sl}^7 k_F^\lambda \tilde{k}_F^\lambda}
\Big(\frac{m^\ast v_{sl}}{\hbar \tilde{k}_F^\lambda}\Big)^3 (k_B T)^5,
\end{eqnarray}

\begin{eqnarray}\label{apprx_PEl}
\Big \la \frac{1}{\tau^\lambda}\Big \ra_{\rm PE, l} \simeq 
\frac{9(eh_{14})^2 \tilde{g}_\alpha^\lambda 3! \zeta(3)}{2\pi \rho_m \hbar^4 v_{sl}^5 k_F^\lambda \tilde{k}_F^\lambda}
\Big(\frac{m^\ast v_{sl}}{\hbar \tilde{k}_F^\lambda}\Big)^5 (k_B T)^3,
\end{eqnarray}

and

\begin{eqnarray}\label{apprx_PEt}
\Big \la \frac{1}{\tau^\lambda}\Big\ra_{\rm PE, t} &\simeq& 
\frac{(eh_{14})^2\tilde{g}_\alpha^\lambda 3! \zeta(3)}{4\pi \rho_m \hbar^4 v_{st}^5 k_F^\lambda \tilde{k}_F^\lambda}
\Big(\frac{m^\ast v_{st}}{\hbar \tilde{k}_F^\lambda}\Big)^3 (k_B T)^3  \nonumber\\
&\times& \Big\{1+8\Big(\frac{m^\ast v_{st}}{\hbar \tilde{k}_F^\lambda}\Big)^4\Big\}.
\end{eqnarray}

In deriving approximate analytical results for phonon-drag thermopower
we expand velocity in Eq. (\ref{vel}) and retain terms up to $q$ since
$q\ll k_F$ in the BG regime. We find the approximate expression for the 
following quantity as
\begin{eqnarray}\label{vel_apprx}
 v_{k_F+q}^\lambda - v_{k_F}^\lambda
=\Big(\frac{\hbar}{m^\ast}+\lambda
\frac{\tilde{\alpha}^3}{\hbar\alpha\Delta}\Big)q.
\end{eqnarray}

Inserting Eq. (\ref{vel_apprx}) in Eq. (\ref{phdrag})
and doing the integration over $q$ we arrive at the following approximate result for
the phonon-drag thermopower
\begin{eqnarray}\label{pdrag}
S_g^\lambda&\simeq&-\frac{{m^\ast}^2 l_p v_s N_F^\lambda}{4\pi^2 n e \hbar^2 k_F^\lambda k_B T^2}
\Big(\frac{m^\ast v_{s}}{\hbar \tilde{k}_F^\lambda}\Big)^3 \Big(\frac{\hbar}{m^\ast}+\lambda
\frac{\tilde{\alpha}^3}{\hbar\alpha\Delta}\Big)\nonumber\\
&\times& \int dq_\bot q_\bot^4 \vert C_{q_\bot} \vert^2 N_{q_\bot}(N_{q_\bot}+1).
\end{eqnarray}

After doing the integration over $q_\bot$ we finally obtain the following expressions for the
phonon-drag thermopower due to DP, longitudinal, and transverse PE scatterings, respectively
\begin{eqnarray}\label{Sg_DP}
S_{g}^\lambda \Big\vert_{\rm DP}\simeq-\frac{k_B}{e} \frac{D^2}{\hbar^2v_{sl}^2}
\Big(\frac{m^\ast v_{sl}}{\hbar \tilde{k}_F^\lambda}\Big)^3
P_l^\lambda   5! \zeta(5) (k_B T)^4,
\end{eqnarray}

\begin{eqnarray}\label{Sg_PEl}
S_{g}^{\lambda}\Big\vert_{\rm PE, l} \simeq-\frac{k_B}{e} \frac{9}{2}
\Big(\frac{m^\ast v_{sl}}{\hbar \tilde{k}_F^\lambda}\Big)^5
(eh_{14})^2 P_l^\lambda 3! \zeta(3) (k_B T)^2,
\end{eqnarray}

and

\begin{eqnarray}\label{Sg_PEt}
S_{g}^\lambda \Big\vert_{\rm PE, t}&\simeq&-\frac{k_B}{e} \frac{1}{4}\frac{v_{st}}{v_{sl}}
\Big(\frac{m^\ast v_{st}}{\hbar \tilde{k}_F^\lambda}\Big)^3
\Big\{1+8\Big(\frac{m^\ast v_{st}}{\hbar \tilde{k}_F^\lambda}\Big)^4\Big\}\nonumber\\
&\times& (eh_{14})^2 P_t^\lambda   3! \zeta(3) (k_B T)^2.
\end{eqnarray}
Here $P_{l(t)}^\lambda$ is defined as 
\begin{eqnarray}
P_{l(t)}^\lambda=\frac{m^{\ast^2}l_pN_F^\lambda}{8\pi^2n\rho_m\hbar^5v_{sl(t)}^4 k_F^\lambda}
\Big(\frac{\hbar}{m^\ast}+\lambda\frac{\tilde{\alpha}^3}{\hbar\alpha\Delta}\Big).
\end{eqnarray}

Let us now provide here a systematic comparison between the results obtained for a QW (present case) and
two-dimensional electron system (2DES) with Rashba SOI in BG regime. In this context, we calculate
the following quantity $S_g\rho^{-1}$ for both quasi-2DES and QW. In the case of a quasi-2DES,\cite{tutul1,tutul2}
using approximate analytical expressions for $\rho$ and $S_g$ in the BG regime, one can 
obtain the following result: $S_g\rho^{-1}\big\vert_{2d}\simeq -\Gamma k_{f0}^2\kappa_{2d}^2/T$ for DP and 
longitudinal PE scattering. Here, $\Gamma=el_pv_{sl}/(4\pi)$, $\kappa_{2d}=1-2k_{\alpha}^2/k_{f0}^2$ and 
finally, $k_{f0}$ is the Fermi wave vector obtained via $k_{f0}=\sqrt{2\pi n_{2d}}$ with $n_{2d}$ as the
carrier concentration in 2D. The result corresponding to the transverse PE scattering is easily
obtained by multiplying the above result by a factor of $v_{st}^2/v_{sl}^2$. Note that  $\rho$ ($S_g$)
represents the total resistivity (phonon-drag thermopower) which is obtained by summing up the contributions
coming from individual energy branches. In the present case, with $B\neq 0$, the total resistivity or
phonon-drag thermopower can not be obtained because of the complicated structure of $k_F^\lambda$ and
other quantities as evident from Eqs.~\eqref{apprx_DP}-\eqref{apprx_PEt} and Eqs.~\eqref{Sg_DP}-\eqref{Sg_PEt}. However,
in the $B\rightarrow 0$ limit, total quantities can be obtained easily. In this limit, we obtain
$S_g\rho^{-1}\big\vert_{1d}\simeq -\Gamma k_F^0\kappa_{1d}/(2T)$ due to DP and longitudinal PE scattering, where
$\kappa_{1d}=1-(k_{\alpha}/k_F^0)^2$ and $k_F^0=n\pi/2$ with $n$ as the electron density in 1D. 
Multiplying $S_g\rho^{-1}\big\vert_{1d}$ by $v_{st}^2/v_{sl}^2$, one can obtain the corresponding 
result for transverse PE case. However, the functional forms of $S_g\rho^{-1}\big\vert_{1d}$ and
$S_g\rho^{-1}\big\vert_{2d}$ are different but in both cases we essentially obtain $S_g\rho^{-1}\sim T^{-1}$
which confirms Herring's law.\cite{herring}

\section{Results and Discussions}
From Eqs. (\ref{apprx_DP}-\ref{apprx_PEt}) and Eqs. (\ref{Sg_DP}-\ref{Sg_PEt}) 
it is revealed that phonon limited resistivity and phonon-drag thermopower 
in the BG regime depends on temperature in a power law fashion i.e. we have 
$\rho$ or $S_g\sim T^{\nu_T}$. 
The effective exponent $\nu_T$ becomes $5$ and 
$3$ due to DP and PE scattering, respectively in the case of resistivity. 
On the other hand, for phonon-drag thermopower $\nu_T$ is $4$ and $2$ corresponding
to DP and PE scattering, respectively. However, the integrals over $q_\bot$ in Eqs. (\ref{IRT_fn}) and (\ref{phdrag}) 
have been evaluated numerically for both DP and PE scattering mechanisms to the 
show the explicit temperature dependence of $\rho$ and $S_g$. 
 
\begin{figure}[h!]
\begin{center}\leavevmode
\includegraphics[width=130mm,height=70mm]{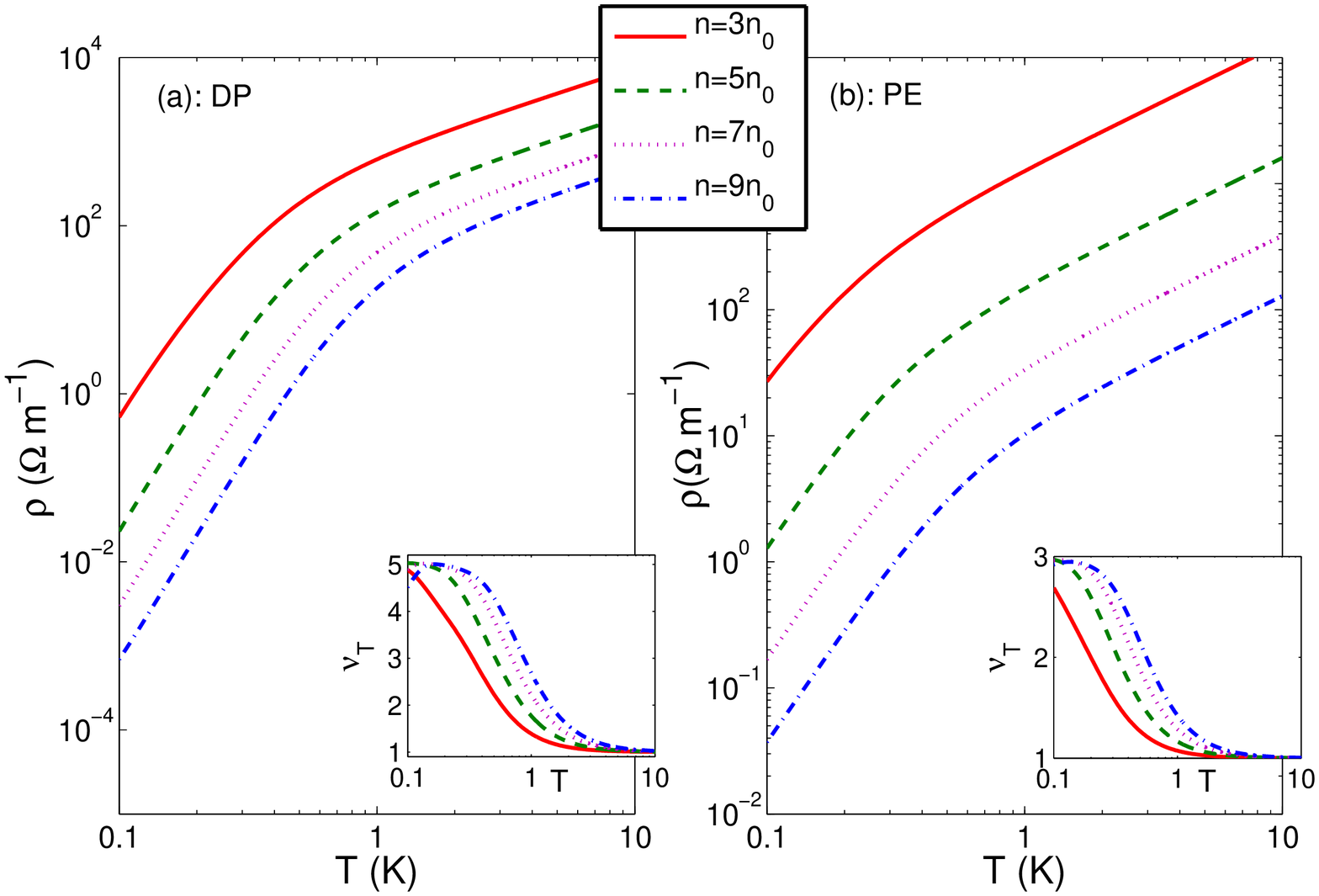}
\caption{(Color online) The dependence of phonon limited resistivity
on temperature is shown. Panel (a) and (b) represent DP and PE scattering contribution,
respectively. Different values of density namely $n=3n_0$, $5n_0$, $7n_0$, and $9n_0$
are considered. The strength of RSOI is fixed to $\alpha=3\alpha_0$.
The temperature variations of the exponent $\nu_T$ are shown in 
the insets of both panels.}
\label{Fig1}
\end{center}
\end{figure} 
 
\begin{figure}[h!]
\begin{center}\leavevmode
\includegraphics[width=130mm, height=70mm]{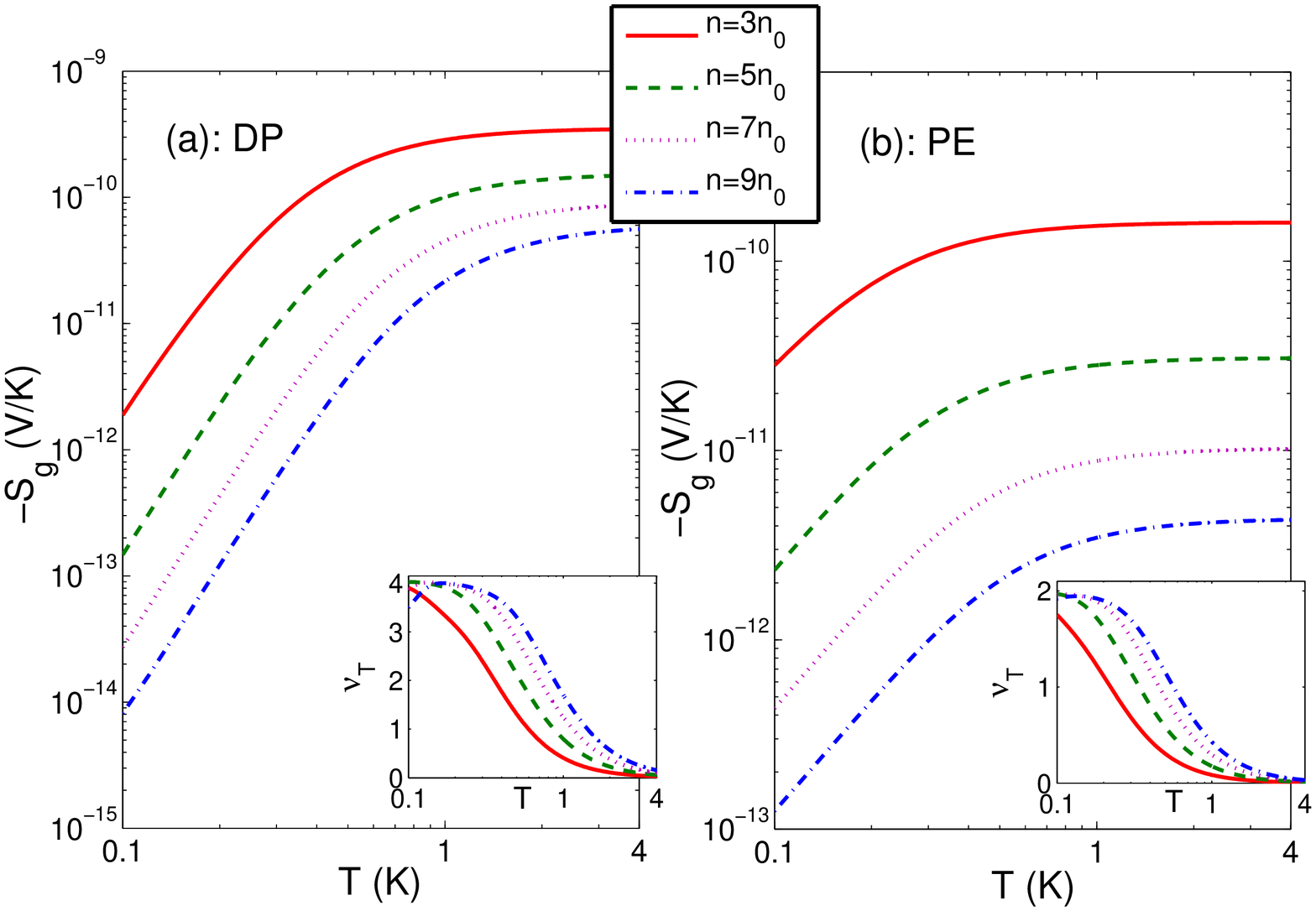}
\caption{(Color online)  Temperature dependence of phonon-drag thermopower due 
to DP and PE scatterings are shown. Different values of density namely $n=3n_0$, $5n_0$, $7n_0$, and $9n_0$
are considered. The strength of RSOI is fixed to $\alpha=3\alpha_0$. The temperature 
dependencies of the effective exponent of phonon-drag thermopower are shown in the 
insets of both panels.}
\label{Fig2}
\end{center}
\end{figure}

For the numerical calculation following 
material parameters, appropriate for an InAs quantum wire, have been considered: 
$m^\ast=0.036m_e$ with free electron mass $m_e$, $g^\ast=-8$,
$\rho_m=5.68\times10^3$ Kg m$^{-3}$,
$v_{sl}=4.41 \times 10^3$ ms$^{-1}$, 
$v_{st}=2.35\times10^3$ ms$^{-1}$,
$D=-5.08$ eV, $h_{14}=3.5\times10^8$ Vm$^{-1}$, 
$n_0=10^{7}$ m$^{-1}$, $R_0=10$ nm, and
$\alpha_0=10^{-11}$ eVm. 
The value of the external magnetic field is taken to be 
$B=0.3$ T.

Fig. 1 shows the temperature variation of phonon limited resistivity 
for different densities namely $n=3n_0, 5n_0, 7n_0$ and $9n_0$. The 
value of Rashba parameter is considered as $\alpha=3\alpha_0$. Fig. 1
clearly demonstrates a crossover from the low temperature BG regime 
to high temperature equipartition regime (in which $\rho \sim T$). 
For both DP and PE scattering mechanisms $\rho$ decreases as $n$ increases.
The resistivity due to PE scattering is higher in magnitude than DP
scattering. The exponent $\nu_T$ of the temperature dependence of $\rho$ 
can be defined as $\nu_T=d{\rm{log}}\rho/d{\rm log}T$ which is extracted numerically
and its variation with temperature has been shown in the insets of Fig. 1. It is clear
that the temperature variation of $\nu_T$ depends on electron density.
At lower density, namely $n=3n_0$ the exponent $\nu_T$ shows a clear 
deviation from the limiting case (i.e. $\nu_T=5$ due to DP and $\nu_T=3$ for 
PE scattering). As density increases the BG temperature regime becomes more 
stable. This numerically obtained BG regime is in excellent agreement
with the approximated analytical results. As temperature increases $\nu_T$
approaches to its equipartition value i.e. $\nu_T=1$.
 
\begin{figure}[h]
\begin{center}\leavevmode
\includegraphics[width=120mm, height=60mm]{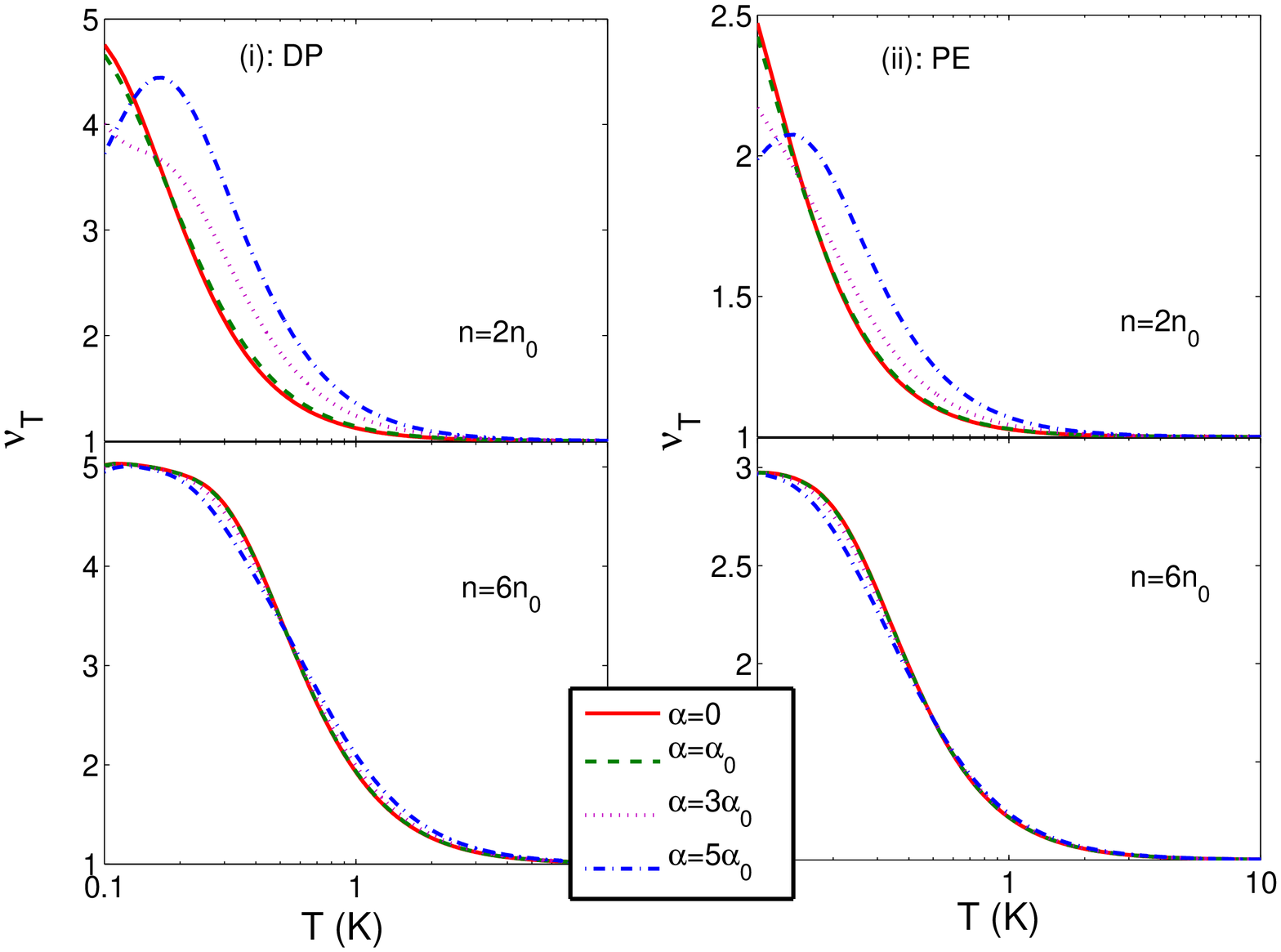}
\caption{(Color online) The temperature variation of the effective exponent of
resistivity i.e. $\nu_T=d{\rm log}\rho/d{\rm log}T$ for various values of $\alpha$, namely, $\alpha=0$, $\alpha_0$,
$3\alpha_0$ and $5\alpha_0$ are shown. Left panel is considered for DP scattering
in which upper and lower panel correspond to $n=2n_0$ and $6n_0$. Similarly right panel 
describes PE scattering for $n=2n_0$ and $6n_0$.}
\label{Fig3}
\end{center}
\end{figure}

\begin{figure}[h]
\begin{center}\leavevmode
\includegraphics[width=120mm, height=60mm]{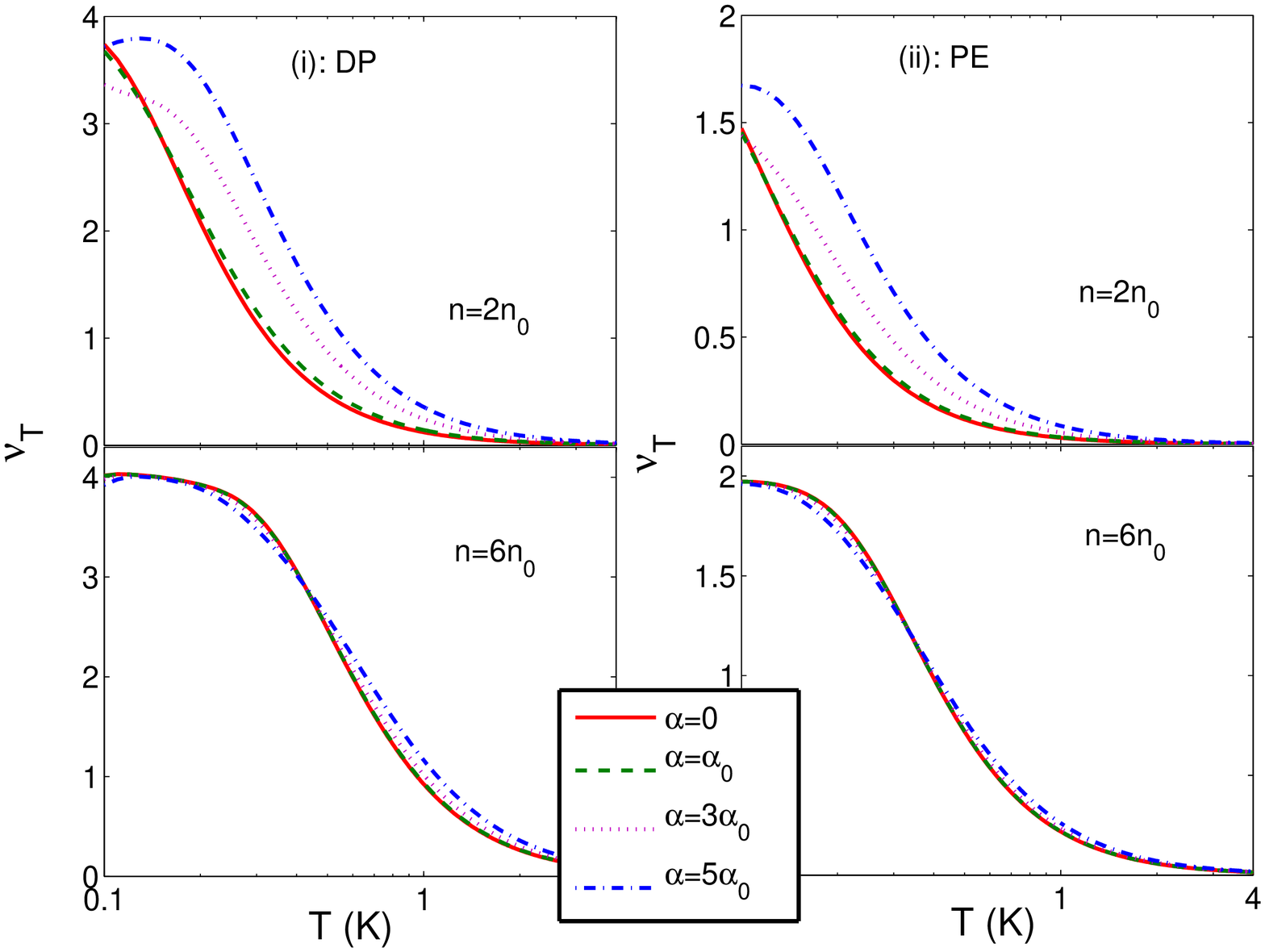}
\caption{(Color online) The temperature variation of the effective exponent 
of phonon-drag thermopower i.e. $\nu_T=d{\rm log} S_g/d{\rm log}T$ for various values of $\alpha$, 
namely, $\alpha=0$, $\alpha_0$, $3\alpha_0$ and $5\alpha_0$ are shown. Left panel is considered for DP scattering
in which upper and lower panel correspond to $n=2n_0$ and $6n_0$. Similarly right panel 
describes PE scattering for $n=2n_0$ and $6n_0$.}
\label{Fig4}
\end{center}
\end{figure}

In Fig. 2 we show the temperature dependence of phonon-drag thermopower
due to DP and PE scattering. $S_g$ decreases with the increase of density.
In this case we also extract the exponent $\nu_T=d{\rm log}S_g/d{\rm log}T$ of 
the temperature dependence of $S_g$. Similar to the resistivity case the 
temperature dependence of $\nu_T$ also depends on the density as depicted 
in the insets. At higher density BG regime is obtained in which $S_g\sim T^4$
due to DP and $S_g\sim T^2$ due to PE scattering. The magnitude of $S_g$ due 
PE scattering is higher than that of DP scattering.

\begin{figure}[h!]
\begin{center}\leavevmode
\includegraphics[width=160mm,height=70mm]{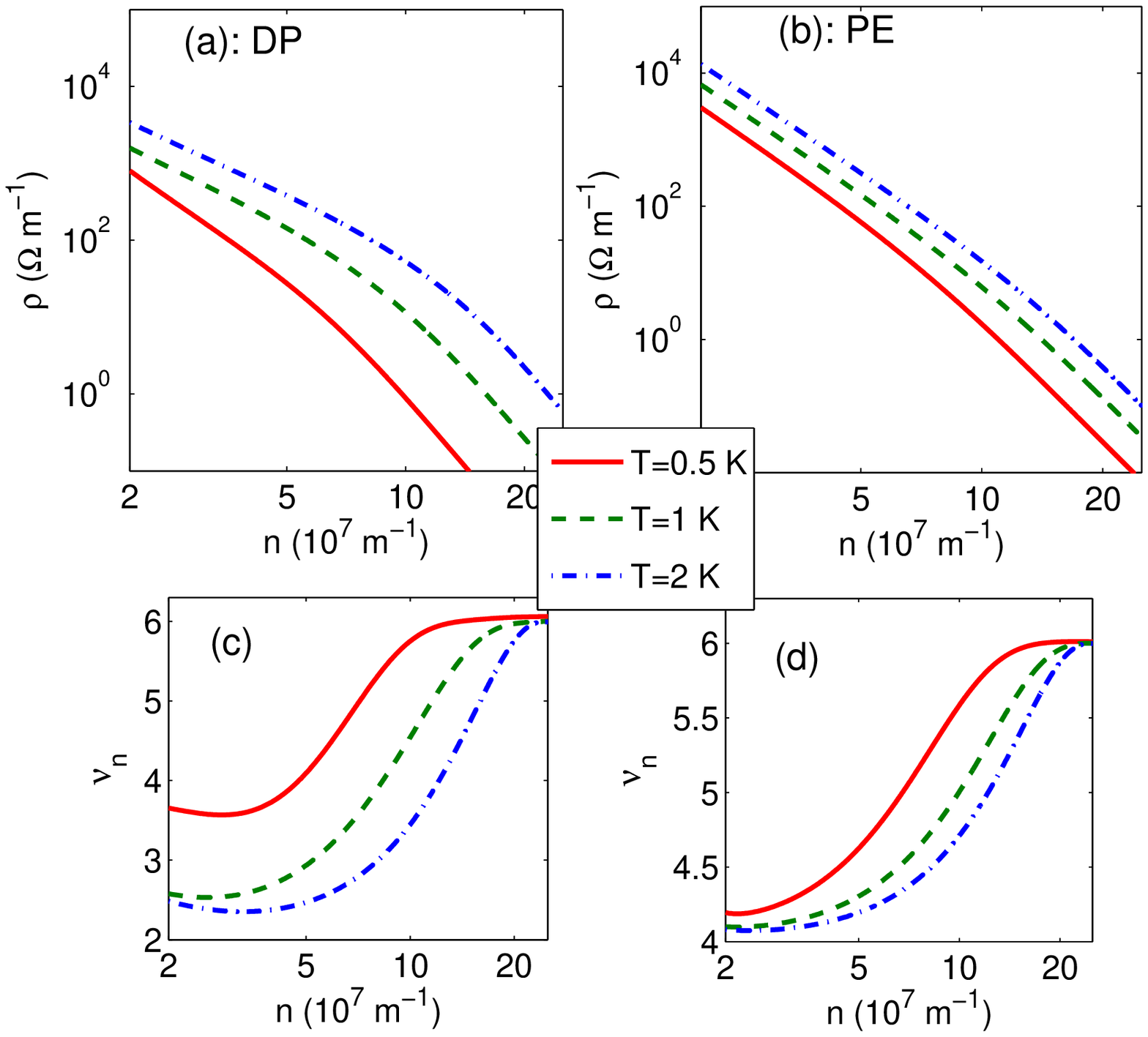}
\caption{(Color online) The dependence of phonon limited resistivity
on carrier density for different temperatures namely, $T=0.5$ K, $1$ K, and 
$2$ K are shown. We consider $\alpha=2\alpha_0$. Panels (a) and (b) are due to 
DP and PE scattering. In panels (c) and (d) we show the variation of 
the quantity $\nu_n=-d{\rm log}\rho/d{\rm log}n$ with $n$ due to DP and 
PE scattering respectively.}
\label{Fig5}
 \end{center}
\end{figure}

\begin{figure}[h!]
\begin{center}\leavevmode
\includegraphics[width=160mm,height=70mm]{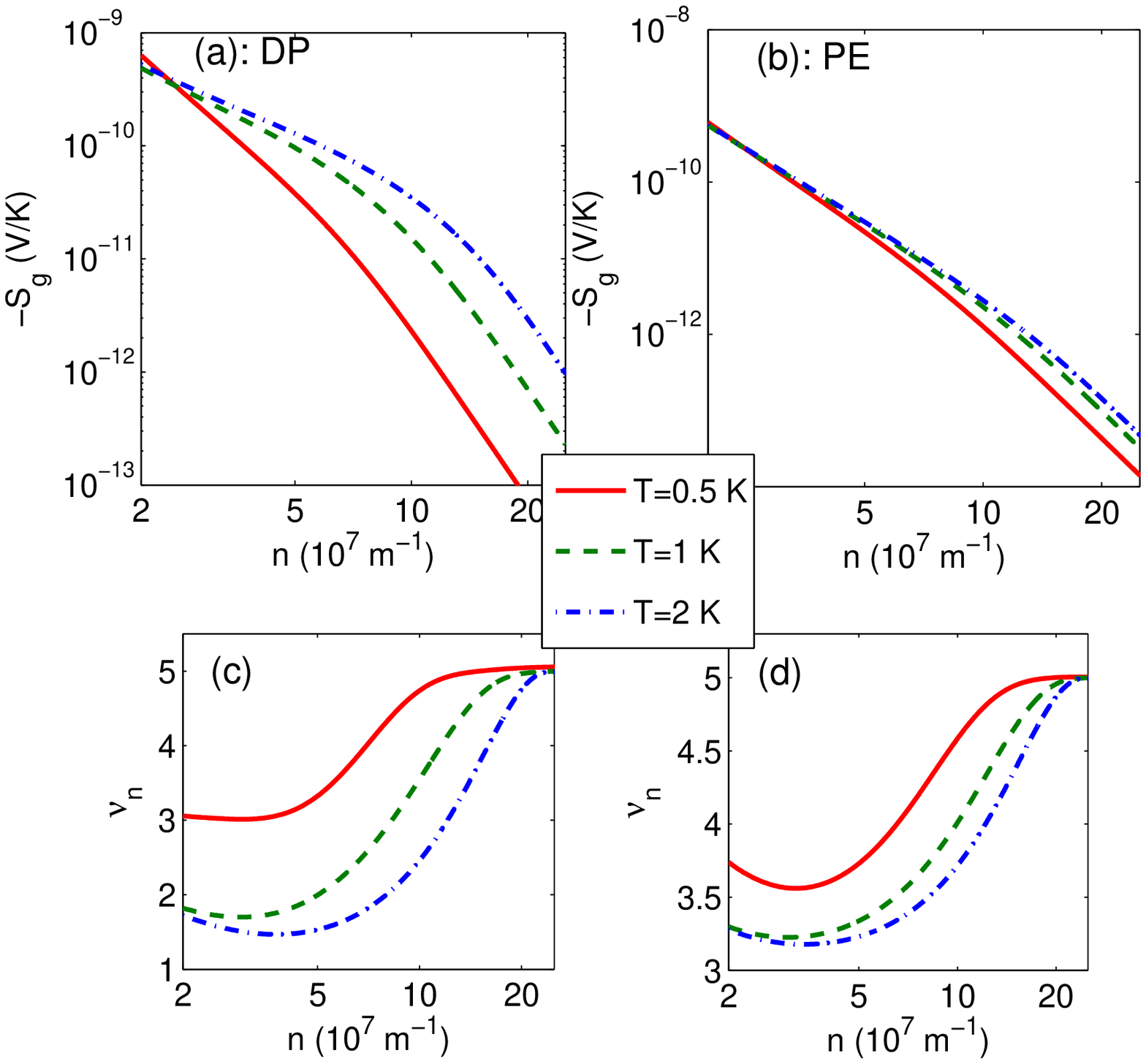}
\caption{(Color online) The dependence of phonon-drag thermopower
on carrier density for different temperatures namely, $T=0.5$ K, $1$ K, and 
$2$ K are shown. We consider $\alpha=2\alpha_0$. Panels (a) and (b) are due to 
DP and PE scattering. In panels (c) and (d) we show the variation of 
the quantity $\nu_n=-d {\rm log}S_g/d{\rm log} n$ with $n$ due to DP and 
PE scattering respectively.}
\label{Fig7}
\end{center}
\end{figure}

Let us now discuss the following important point. The boundary 
of the BG regime is defined by the characteristic temperature
$T_{BG}=2\hbar v_s k_F/k_B$. For a typical value of electron density,
say $n=5n_0$ we have $T_{BG}\sim 6$ K. But it is obtained numerically
that the BG regime exists for a small range of temperature below $1$ K.

The temperature dependence of $\nu_T$ not only depends on the density
but also on the Rashba parameter $\alpha$. These facts are depicted in 
Figs. 3 and 4 in which the temperature dependencies of $\nu_T$ corresponding to 
$\rho$ and $S_g$ for different $\alpha$ are shown. When density is low the temperature 
variation depends significantly on $\alpha$. At a relatively higher density,
the effect of $\alpha$ on this temperature dependence is not so prominent for both $\rho$ and $S_g$. 
Similar effect of $\alpha$ on the temperature dependence of $\rho$ or 
$S_g$ in a Rashba spin-orbit coupled two dimensional electron gas in 
BG regime has been addressed recently.\cite{tutul1,tutul2}

In Figs. (5) and (6) we have shown how $\rho$ and $S_g$ depend on the 
electron density at a fixed temperature in BG regime.
In Eqs. (\ref{apprx_DP}-\ref{apprx_PEt}) and (\ref{Sg_DP}-\ref{Sg_PEt}) 
one can notice that both IRT (and consequently $\rho$) and phonon-drag thermopower
show a power law dependence with electron density through the Fermi wave vectors
at a fixed temperature. So in general we can write $\rho$ or $S_g\sim n^{-\nu_n}$,
where the exponent $\nu_n$ corresponding to $\rho$ and $S_g$ can be obtained by 
taking negative logarithmic differentiation of $\rho$ or $S_g$ with respect to $n$
i.e. $\nu_n=-d{\rm log}\rho(S_g)/d{\rm log}n$. Let us now estimate $\nu_n$ from
Eqs. (\ref{apprx_DP}-\ref{apprx_PEt}) and (\ref{Sg_DP}-\ref{Sg_PEt}). It is 
well known that the Fermi wave vector scales with density as $k_F^0\sim n$ in one dimension.
In our case $k_F^\lambda$ depends on density in a complicated way as seen from 
Eq.~\eqref{ferm_wave}. Nevertheless, we can find $k_F^\lambda \sim n$ since 
$k_\alpha, \Delta/\alpha\ll 2m^\ast\epsilon_F/\hbar^2$.
From Eqs. (\ref{apprx_DP}-\ref{apprx_PEt})
one finds $\la \tau^{-1}\ra\sim n^{-5}$ and as a result we have $\rho \sim n^{-6}$.
The phonon-drag thermopower depends on density as $S_g\sim n^{-5}$ as seen from
Eqs. (\ref{Sg_DP}-\ref{Sg_PEt}). However solving Eqs. (\ref{IRT_fn}) and (\ref{phdrag})
numerically we find that $\nu_n$ undergoes a crossover from a relatively lower density behavior to 
a higher density behavior for both $\rho$ and $S_g$. As density increases $\nu_n$ approaches towards 
the values obtained from asymptotic expressions i.e. $\nu_n=6$ for $\rho$ and $\nu_n=5$ for
$S_g$. Note that at higher density same values of $\nu_n$ are obtained due to DP and PE scattering
for both case of $\rho$ and $S_g$. But at lower densities $\nu_n$ differs significantly 
due to DP and PE scattering.

Although a gap $\Delta$ is considered in the energy spectrum but its magnitude 
is much smaller than that corresponding to the Rashba spin-splitting
i.e. $\Delta \ll \alpha k_F$. The main purpose for considering $\Delta$ is to 
see whether inter-branch transitions are happening or not. But in the 
BG regime, the possibility of inter-branch scattering has been ruled out.
So the qualitative results do not change significantly due to the presence
of $\Delta$ in the energy spectrum.

\section{Summary} 
In summary we have studied various features of acoustic phonon limited
resistivity and phonon-drag thermopower in a Rashba spin-orbit 
coupled semiconductor QW with an in-plane magnetic field induced gap.
Two mechanisms of electron-phonon interaction, 
namely, DP and PE scatterings are taken into consideration.
In the BG regime a power law dependence of both resistivity and 
phonon-drag thermopower with temperature have been obtained analytically. We find the 
exponent ($\nu_T$) of the temperature dependence which takes the value $5$ and $3$ corresponding 
to the DP and PE scattering, respectively in the case of resistivity. 
$\nu_T$ becomes $4$ and $2$ in the case of phonon-drag thermopower due to 
DP and PE scattering, respectively. Through a numerical calculation, we have 
shown a transition in resistivity from BG to equipartition regime. Numerically,
it is also found that $\nu_T$ depends on both density and Rashba parameter.
At higher density $\nu_T$ matches well with that obtained from the analytical 
calculation for both $\rho$ and $S_g$ or in other words a BG regime is established
at higher density. The effect of spin-orbit interaction on $\nu_T$ is found
to be more prominent in low density regime. Finally the dependence of $\rho$
and $S_g$ on the carrier density are also discussed. An approximate 
analytical calculation shows that $\rho\sim n^{-6}$ and $S_g\sim n^{-5}$ in
the BG regime. These dependence on $n$ have been confirmed through a numerical analysis
at higher densities. The results obtained in the present case have also been
compared with the corresponding results for spin-orbit coupled two-dimensional
electron system and we obtain in both cases $S_g\rho^{-1}\sim T^{-1}$ which affirms
Herring's law.

\appendix
\section{}
In this Appendix we shall perform an explicit derivation 
of the term $\delta(\epsilon_{k_F\pm q}^\lambda-\epsilon_{k_F}^\lambda \mp \hbar\omega_Q)$ 
as given in Eq. (\ref{delta_fn}.)

From Eq. (\ref{eig_eng}) one can write
\begin{eqnarray}\label{eig_app}
\epsilon_{k_F+q}^\lambda=\frac{\hbar^2 (k_F^\lambda+q)^2}{2m^\ast} + \lambda
\sqrt{\alpha^2 (k_F^\lambda+q)^2+\Delta^2}.
\end{eqnarray}

Now defining $\varepsilon_{k_F^\lambda}=\sqrt{(\alpha k_F^\lambda)^2+\Delta^2}$
and assuming $q\ll k_F^\lambda$, the 
second term in Eq. (\ref{eig_app}) can be expanded up to $q^2$ as
\begin{eqnarray}
\varepsilon_{{k_F^\lambda}+q}=\varepsilon_{k_F^\lambda}+\frac{\alpha^2 k_F^\lambda q}{\varepsilon_{k_F^\lambda}}
+\frac{\alpha^2 q^2}{2\varepsilon_{k_F^\lambda}}\Big(1-\frac{\alpha^2 k_F^{\lambda^2}}{\varepsilon_{k_F^\lambda}^2}\Big).
\end{eqnarray}

We then have
\begin{eqnarray}\label{del_apprx}
\epsilon_{k_F+ q}^\lambda-\epsilon_{k_F}^\lambda=\frac{\hbar^2}{2m^\ast}
\tilde{g}_\alpha^\lambda \Big(q^2 
+ 2q\frac{\tilde{k}_F^\lambda}{\tilde{g}_\alpha^\lambda}\Big),
\end{eqnarray}
where $\tilde{g}_\alpha^\lambda$ and $\tilde{k}_F^\lambda$ are defined earlier.

Since we are dealing with the BG regime in which $q\ll k_F^\lambda$, the 
term $q^2$ in Eq. (\ref{del_apprx}) can be neglected. So from the 
energy conservation $\epsilon_{k_F\pm q}^\lambda-\epsilon_{k_F}^\lambda \mp \hbar\omega_Q=0$,
one can obtain 
$q=\big(m^\ast v_s/\hbar \tilde{k}_F^\lambda\big)Q$
with $Q=\sqrt{q_\bot^2+q^2}$. 
Since the coefficient $m^\ast v_s/(\hbar \tilde{k}_F^\lambda)\ll 1$ and consequently
we have $q\ll Q$ which in turn forces us to write the following expression
\begin{eqnarray}
q=\frac{m^\ast v_s}{\hbar \tilde{k}_F^\lambda}q_\bot.
\end{eqnarray}

We now calculate the delta function corresponding to the absorption case
which can be obtained in the following form
\begin{eqnarray}\label{delta2}
\delta(\epsilon_{k_F+ q}^\lambda&-&\epsilon_{k_F}^\lambda - \hbar\omega_Q)
=\frac{2m^\ast}{\hbar^2 \tilde{g}_\alpha^\lambda}
\delta \Big(q^2 + 2q\frac{\tilde{k}_F^\lambda}{\tilde{g}_\alpha^\lambda}
- \frac{2m^\ast v_s q_\bot}{\hbar \tilde{g}_\alpha^\lambda}\Big)\nonumber\\
&=&\frac{2m^\ast}{\hbar^2 \tilde{g}_\alpha^\lambda}\frac{1}{\vert q_+-q_-\vert}
\Big\{\delta(q-q_+)+\delta(q-q_-)\Big\},
\end{eqnarray}
where $q_\pm=\Big(-\tilde{k}_F^\lambda+\sqrt{\tilde{k}_F^{\lambda^2}+C^\lambda q_\bot}\Big)/\tilde{g}_\alpha^\lambda$
with $C^\lambda=2m^\ast v_s \tilde{g}_\alpha^\lambda/\hbar$. With the approximation
$q\ll k_F^\lambda$ one can find $q_+=m^\ast v_s q_\bot/(\hbar \tilde{k}_F^\lambda)$
and $q_-=-2\tilde{k}_F^\lambda/\tilde{g}_\alpha^\lambda - m^\ast v_s q_\bot/(\hbar \tilde{k}_F^\lambda)$.
Since we are considering BG regime then one may ignore the term $\delta(q-q_-)$ in Eq. (\ref{delta2}).
Exactly similar analysis can be done for emission case. Now it is straightforward to obtain Eq. (\ref{delta_fn})
from Eq. (\ref{delta2}).

\end{document}